\documentclass[10pt,aps,pre,showpacs,twocolumn]{revtex4-1}

\usepackage{graphicx}
\usepackage{amsmath}
\usepackage[english]{babel}

\begin{document}


\title{Towards a Fluctuation Theorem in an Atmospheric Circulation Model}


\author{B. Schalge}
\affiliation{Meteorologisches Institut, KlimaCampus, 
Universit\"at Hamburg, Hamburg, Germany}
\author{R. Blender}
\author{J. Wouters}
\author{K. Fraedrich}
\author{F. Lunkeit}
\affiliation{Meteorologisches Institut, KlimaCampus, 
Universit\"at Hamburg, Hamburg, Germany}


\date{\today}

\begin{abstract}
An investigation of the distribution of finite time trajectory divergence is performed on
an Atmospheric Global Circulation Model. 
The distribution of the largest local Lyapunov exponent shows a significant probability 
for negative values over time spans up to 10 days. This effect is present for resolutions up to 
wave numbers $\ell=42$ ($\approx$ 250km). The probability for a negative local largest Lyapunov exponent decreases over time,
similarly to the predictions of the Fluctuation Theorem for entropy production.
The model used is hydrostatic with variable numbers of 
vertical levels and different horizontal resolutions.
\end{abstract}

{\center revised version}


\maketitle

\section{Introduction}  \label{Sec_Intro}

%



The fluctuation theorem (FT) \cite{evans93,gallavotti95}
describes the probability of negative entropy production
in nonlinear nonequilibrium systems. 
It shows how such a violation of the second law becomes
exponentially unlikely when the period over which entropy production is measured increases.
This FT has been observed to hold in experiments
and numerical simulations of turbulent flows \cite{ciliberto98,gallavotti04}.

In atmospheric models, similar processes can be observed. In this 
case a violation of the second law can be associated with an increase of  
predictability (return of skill) \cite{anderson94},
indicating that for certain flow configurations an error, seen as a finite
perturbation to the trajectory, decreases over time.


Studies of fluctuation theorems are often directed at mesoscopic systems, such as molecular motors or proteins,
having a moderate number of degrees of freedom.
The numerical modelling of continuous systems is restricted to a 
finite number of degrees of freedom which also has
to be moderate if long time spans have to be covered.
Therefore, we can expect that deviations from the second 
law as described by the FT may occur also in such
numerical simulations.

In this paper we consider simulations with an 
atmospheric Global Circulation Model (GCM). Models of this type form the hydrostatic dynamical 
cores of complex weather and climate models. 
They simulate the stratified large scale flow on a rotating sphere driven
by a temperature relaxation and subject to linear surface friction and hyper-diffusion.
The flow is characterised by quasi-two-dimensional turbulence and Rossby waves
\cite{rhines75}. By changing horizontal and vertical resolutions of the model
the number of degrees of freedom can be modified.

Instead of using the common dynamical system formulation of entropy production as the phase space contraction rate,
we choose a simplified approach. 
The entropy production $\Sigma$ 
is identified 
with the 'entropy-like quantity' (related to the Kolmogorov 
or metric entropy)
$k_n$ \cite{casartelli76,benettin76}
defined as the local error growth 
given by the largest local Lyapunov exponents $\hat{\lambda}$.
In the long time limit the $\hat{\lambda}$ tend to the largest Lyapunov exponent
$\lambda_{max}$.
For a summary on the definitions of entropy 
in dynamical systems, in particular the relationship 
with topological entropy see Young \cite{young03}. Previously, studies of the
distribution of $\hat{\lambda}$ have been performed for simple one-dimensional
dynamical systems, showing that these distributions can in certain cases be Gaussian
due to a central limit theorem~\cite{fujisaka83}, however, in general destinct deviations from Gaussianity are found~\cite{yoshida89}.


%

In Sections \ref{Sec_Model} and 
\ref{Sec_Analysis} the model and the analysis are described.
Section \ref{Sec_Results} includes the 
results for the statistics of $\hat{\lambda}$ 
and the validity
of the FT,
and in Section \ref{Sec_SumCon} we summarize the results
and discuss the conclusions.

\section{Global Circulation Model}  \label{Sec_Model}

In this paper we utilize the Portable University Model of the Atmosphere
(PUMA) \cite{fraedrich05,fraedrich12}, a hydrostatic global atmospheric
model based on the multi-layer primitive equations on the sphere.
Model variables are vorticity, horizontal divergence,
temperature and logarithmic surface pressure. 
The complete equations of PUMA are
\begin{eqnarray}
\partial_t \xi
& = &
s^2
\partial_\lambda  {\cal F}_v
-\partial_\mu {\cal F}_u
-\frac{1}{\tau_f} \zeta
-K\nabla^{8} \zeta
\\  \label{puma:a}
%
\partial_t D
& =& s^2
\partial_\lambda
{\cal F}_u
+\partial_\mu {\cal F}_v
-\nabla^2
  [         \frac{s^2}{2} (U^2 + V^2)
\\  \nonumber
&+& \Phi
 +  
\bar{T} \ln p_s  
]
 - 
\frac{1}{\tau_f} D
-K \nabla^{8} D
\\  \label{puma:b}
\partial_t T'
&=&-s^2
\partial_\lambda
(U T')
-\partial_\mu
(V T')
+D T'
-\dot{\sigma}\frac{\partial T}
{\partial \sigma}
\\  \nonumber
&+&\kappa \frac{T\omega}{p}
+\frac{1}{\tau_c}(T_R-T)
-K\nabla^{8} T'
\\    \label{puma:c}
\partial_t \ln p_s
&=&
-s^2 U
\partial_\lambda
\ln p_s
-V \partial_\mu 
\ln p_s -D
-\frac{\partial \dot{\sigma}}
{\partial\sigma}
\\
\label{puma:d}
%
\frac{\partial \Phi}{\partial \ln \sigma}
&=& -T
\label{puma:e}
\end{eqnarray}
where $\mu=\sin \phi$, $s^2 =1/(1- \mu^2)$. 
$\zeta$ and $\xi$ denote absolute and relative vorticity, 
$D$ is the horizontal divergence 
and $p_s$ the surface pressure. The temperature $T$ is 
divided into a background
state, $\bar{T}$, and an anomaly, $T'$. 
Spherical coordinates are given by
$\lambda$ and $\phi$ for longitude and
latitude respectively, $s^2 =1/(1- \sin^2 \phi)$,
$\Phi$ is the geopotential,
$\kappa$ is the adiabatic coefficient, $\omega$ is 
vertical velocity and $K$ a diffusion
coefficient. We also use the abbreviations
$ U = u~ \cos \phi$ and
$V = v~ \cos \phi$ ($u$,$v$ are the zonal and meridional velocities, resp.),
$
{\cal F}_u=V\zeta- \dot{\sigma} \partial U/ \partial\sigma
-T' \partial \ln p_s/ \partial\lambda
$
and
$
{\cal F}_v=-U\zeta-\dot{\sigma}
\partial V/\partial\sigma
-T's^{-2} \partial \ln p_s/\partial \sin \phi
$.
The vertical coordinate
is divided into equally spaced $\sigma$-levels 
($\sigma=p/p_s$, where $p$ and $p_s$ denote the pressure and
the surface pressure, respectively).

To maintain a stationary state the 
model is driven by a Newtonian cooling formulation towards a constant
temperature profile with an equator-to-pole
gradient (i.e., a term $(T_{R}-T)/\tau_c$ is 
added to the temperature equation, where
$\tau_c$ is the time scale, 
$T$ denotes the actual
model temperature and
$T_R$ refers to the prescribed reference temperature).
Dissipation is given by Rayleigh friction in the
boundary layer
(i.e., terms $-\zeta/\tau_f$ and $-D/\tau_f$ are
added to the equations for
vorticity and divergence, where $\tau_f$ is the friction 
time scale). 
Hyper-diffusion ($\propto \nabla^{8}$)
accounts for subscale processes and numerical stability.

%

The equations are numerically solved using the
spectral transform method \cite{orszag70}: linear terms are 
evaluated in the spectral domain while
nonlinear products are calculated in grid point space.
The horizontal resolution is varied between
total spherical wave number $\ell=15$ and 42
(approx.~$7.5^\circ $ and $2.8^\circ$ on the corresponding grid).
The 
wave numbers are restricted
according to the triangular truncation
which is denoted as 'T$\ell$' with the total wave number $\ell$.
The vertical model resolution is between
$L=5$ and 20 vertical levels.
The resolutions considered here restrict the numbers
of degrees of freedom to values below $10^5$,
i.e. to the range of mesoscopic systems.

The model is integrated by a leap-frog method with
a time step of 30min (15min for $\ell=42$ resolution).
Orography is not specified and no external
variability like annual or daily cycles are imposed.  

Due to the neglect of complex parameterizations
like convection,
the error growth in a dynamical core model
corresponds to the long time growth regime in
complex weather and climate models
which incorporate a rapid error growth
caused by small and fast processes
\cite{harlim05}.

\section{Analysis}  \label{Sec_Analysis}

As a measure for predictability in the model 
the linear stability is determined by the simulation 
of two close nonlinear trajectories.
One of the simulations is denoted as reference trajectory $x(t)$
while the second is considered as a perturbed trajectory $\tilde{x}(t)$
initialized with weak random deviations in the surface pressure.
The distance $\| x(t) - \tilde{x}(t) \| $ between the trajectories is determined 
using an Euclidean metric for all model variables in grid point
space. 
Linear instability of the reference trajectory is assessed 
by maintaining small distances through a regular rescaling of 
the perturbation after a time span  $\tau_0=10$ days
to the initial distance $d_0$ (these intervals are denoted as 
rescaling intervals in the following). 
To collapse the perturbation onto 
the most unstable direction 
the analysis neglects an initial spin up of
10 rescaling intervals in the 200 years trajectories.

\begin{figure}[h]
\includegraphics[scale=0.17]{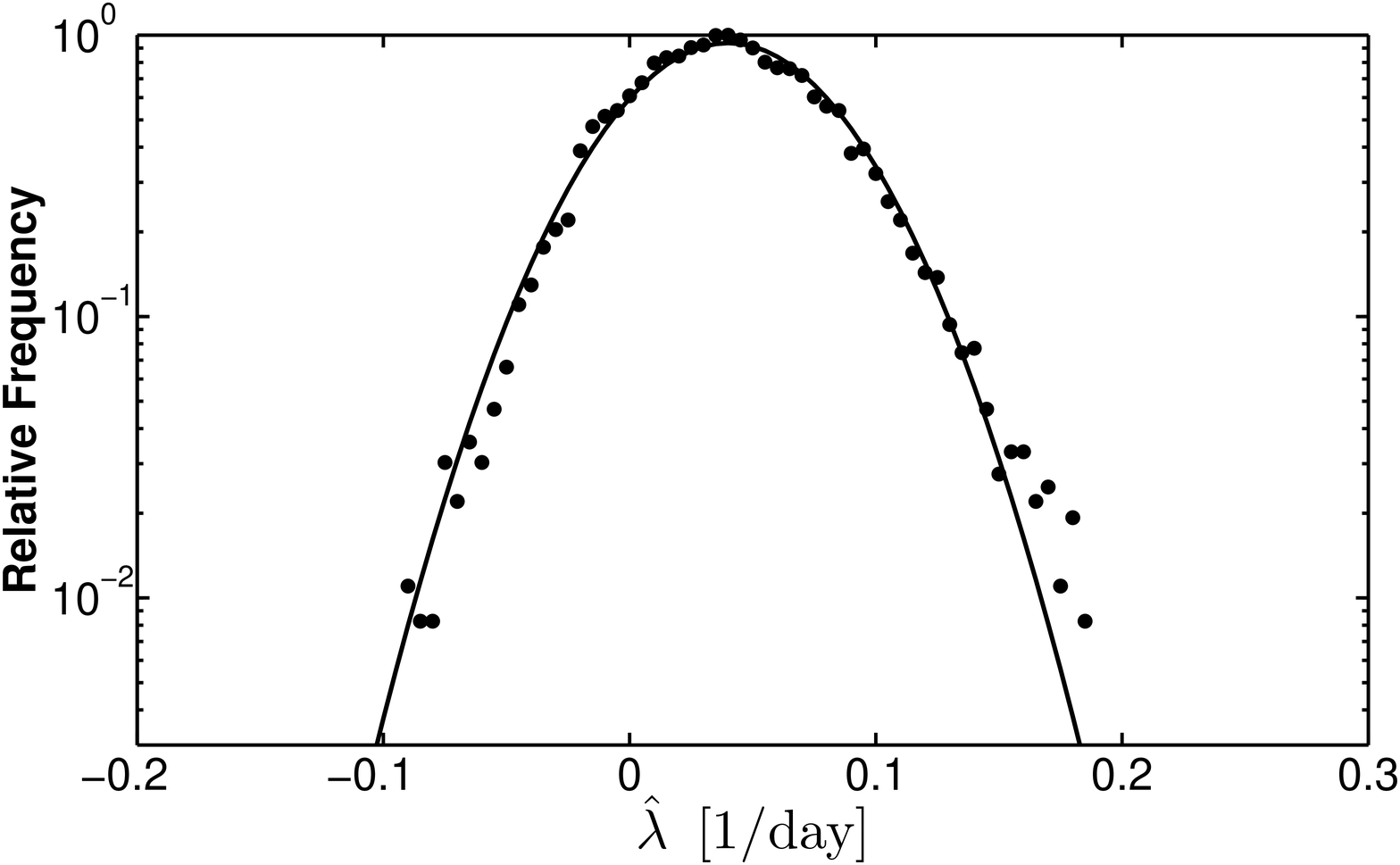}  
\caption{\label{fig1} 
Relative frequencies of the local largest Lyapunov exponent for $\tau=100 \, \text{h}$ with
a Gaussian fit (black) for the resolution T21L5.}
\end{figure}

Within the rescaling intervals of duration $\tau_0$  
the growth of the distance $d$ is measured 
every 5 hours to estimate the 
local largest Lyapunov exponent $\hat{\lambda}$
\begin{equation}
	\hat{\lambda}(t,\tau)
	= \frac{1}{\tau}  \log  \frac{d(t,\tau)}{d_0}
\end{equation}
Here $d(t,\tau) = \| x(t+\tau) - \tilde{x}(t+\tau)\|$,
where $t=n\tau_0$ measures the absolute time and 
$\tau<\tau_0$ is the time elapsed after  
the last rescaling to the initial distance $d_0$. 
The unit for $\hat{\lambda}$ is $1/day$.
In the following the times $t$ characterize the 
rescaling intervals while $\tau$ are the 
growth times within these intervals.

According to \cite{casartelli76,benettin76}
we assess the 
entropy production $\Sigma$ by the  largest 
local Lyapunov exponents.
Periods with $\hat{\lambda}$$<0$ are associated with a 
return of skill in predictability experiment \cite{smith99}.
As the largest exponent is negative, phase space contracts during these intervals, hence the entropy production is negative.

\section{Results}   \label{Sec_Results}

The distribution of $\hat{\lambda}$ is analysed for 
a standard low resolution version and 
for variable vertical and horizontal resolutions. 
It is demonstrated that an approximation of the entropy production 
behaves similarly to the predictions of the Fluctuation Theorem.
This relation holds for 
a range of $\hat{\lambda}$ for a fixed growth time $\tau$
in a low resolution experiment.


In the standard low resolution version T21L5 of the model  
(total wave number 21 and 5 vertical levels) 7744
degrees of freedom are present. 
Due to the numerical efficiency
this resolution is frequently used in long-term simulations
since it produces
a circulation with the characteristic properties
of the observations (for example Hadley and
Ferrel cells and mid-latitude storms \cite{peixoto92}). The exponents are determined for a fixed growth time 
$\tau=100$ hours in all sampling intervals ($\tau_0=10$ days).
Figure \ref{fig1} shows the distribution of the $\hat{\lambda}$ values with a Gaussian fit (normalized to the maximum).
The Gaussian fits in Figure \ref{fig1} and the following figures
are used to characterise the $\hat{\lambda}$ distribution by their means and variances. 
Deviations from Gaussianity cannot be assessed due to the limited
data sets.
About 17\% of the exponents $\hat{\lambda}$ are negative, hence the model
is in a state of negative entropy production for
a considerable amount of time.

\begin{figure}[h]
\includegraphics[scale=0.17]{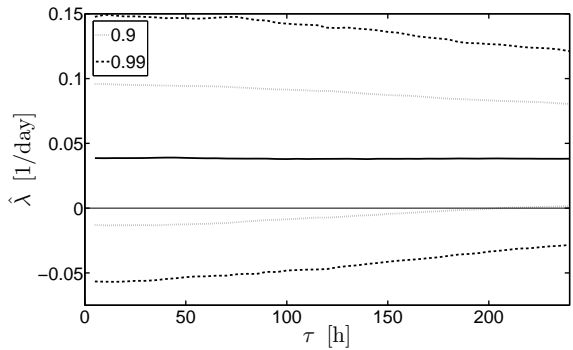}  
\caption{\label{fig2} 
Quantiles (0.9, 0.99) of the local largest Lyapunov exponent.
The median (solid) corresponds to the 
global largest Lyapunov exponent}
\end{figure}


To characterize the distribution for a wide
range of growth times $\tau$ the 0.9 and  0.99-quantiles 
are presented (Figure \ref{fig2}).
The growth times $\tau$ range from the minimum of
5 hours to the maximum given by the rescaling time
$\tau_0=10$ days. Thus  Figure \ref{fig1}
corresponds to the $\tau=100$ hours slice in Figure \ref{fig2}.
The median is the largest Lyapunov exponent
which is by definition independent of the growth time $\tau$.
For increasing $\tau$ the quantiles approach the median
indicating a narrowing distribution.
For large $\tau$ negative values become less likely
and the probability of negative entropy production vanishes.

\begin{figure}[b]
\includegraphics[scale=0.17]{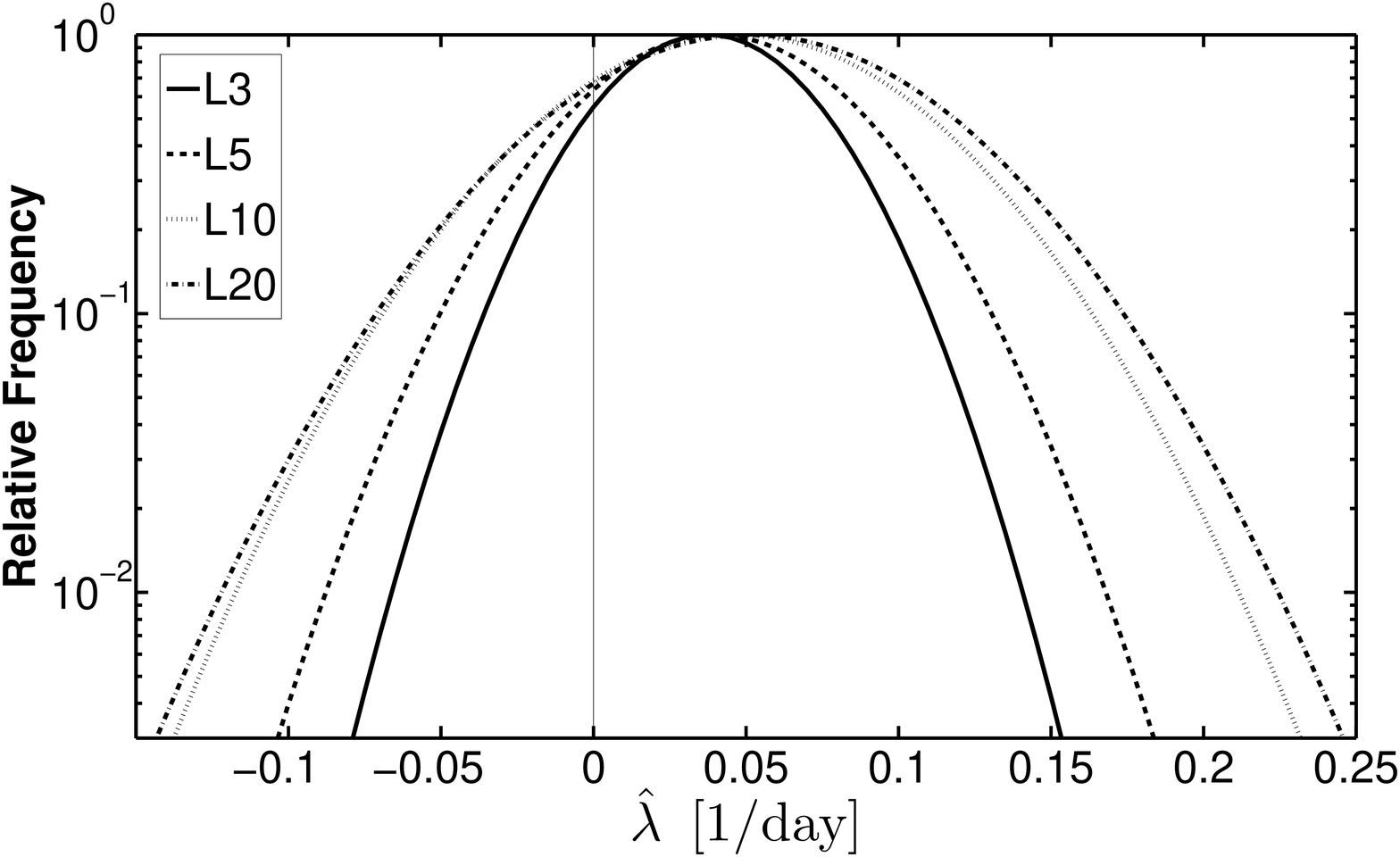}  
\caption{\label{fig3} 
Gaussian fits of the local largest Lyapunov exponent distribution in T21 with different vertical levels $L$
as indicated.
}
\end{figure}


The sensitivity of the $\hat{\lambda}$ distribution 
to the vertical model resolution is determined 
for fixed horizontal resolution T21 (Figure \ref{fig3})
with numbers of levels ranging from 3, 5, 10, to 20.
The number of degrees of freedom 
depends linearly on the number of levels:
4840, 7744, 15004, 29524.
Here Gaussian fits to the distributions 
are shown which are normalized by the maximum
(compare Figure \ref{fig1}).
All curves show a comparable mean but
become broader for increasing numbers of levels.
Beyond 10 levels a limit is reached. 

\begin{figure}[t]
\includegraphics[scale=0.17]{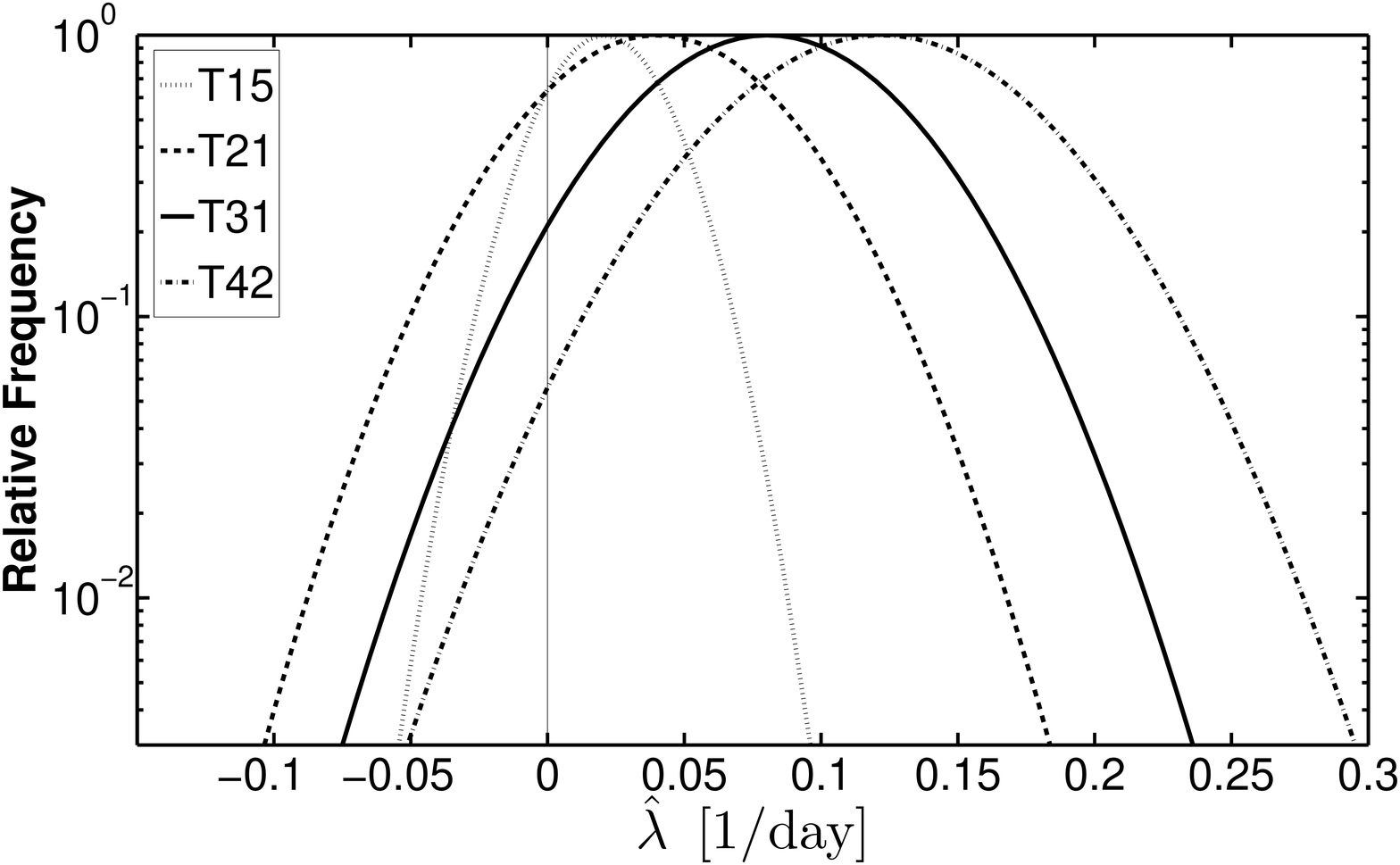}  
\caption{\label{fig4} 
Gaussian fits of the local largest Lyapunov exponent distribution for $L=5$ levels and 
different horizontal resolution.
as indicated.
}
\end{figure}

For increasing horizontal resolutions and
keeping the  number of five levels fixed the $\hat{\lambda}$ distributions
change differently  (Figure \ref{fig4}).
The degrees of freedom for five levels
are 4096 (T15), 7744 (T21), 16384 (T31) and 28224 (T42).

From T15 to T21 the distribution shifts to more
positive means and becomes broader.
Similarly to the analysis for the vertical resolution (Figure \ref{fig3}), an increase in the number of degrees of freedom seems to imply an increase in standard deviation.

From T21 to T31 and T31 to T42 
the main change is a shift to positive values 
while the variance increase is negligible 
(in contrast to Figure \ref{fig3}).
Therefore, the frequency of negative $\hat{\lambda}$ decreases 
with increasing horizontal resolution. 

The increase of the horizontal resolution 
enhances the effective number of degrees of freedom
by the simulation of smaller structures
like mid-latitude vortices.
In contrast,  the increase of the vertical resolution 
beyond 10 levels seems not to
yield additional effective degrees of freedom
for the given horizontal T21 resolution.

\begin{figure}[h]
\includegraphics[scale=0.17]{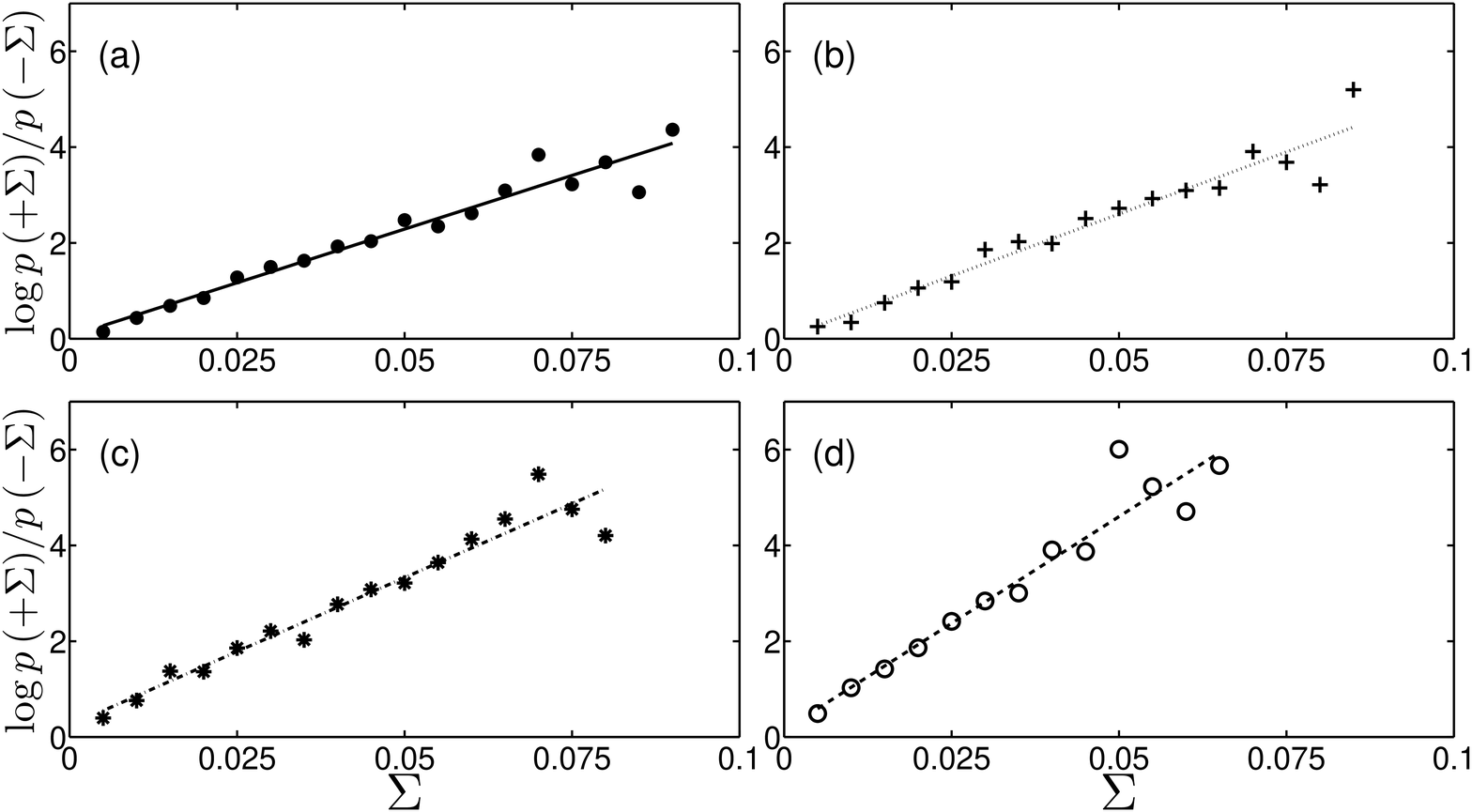}\\  
\includegraphics[scale=0.17]{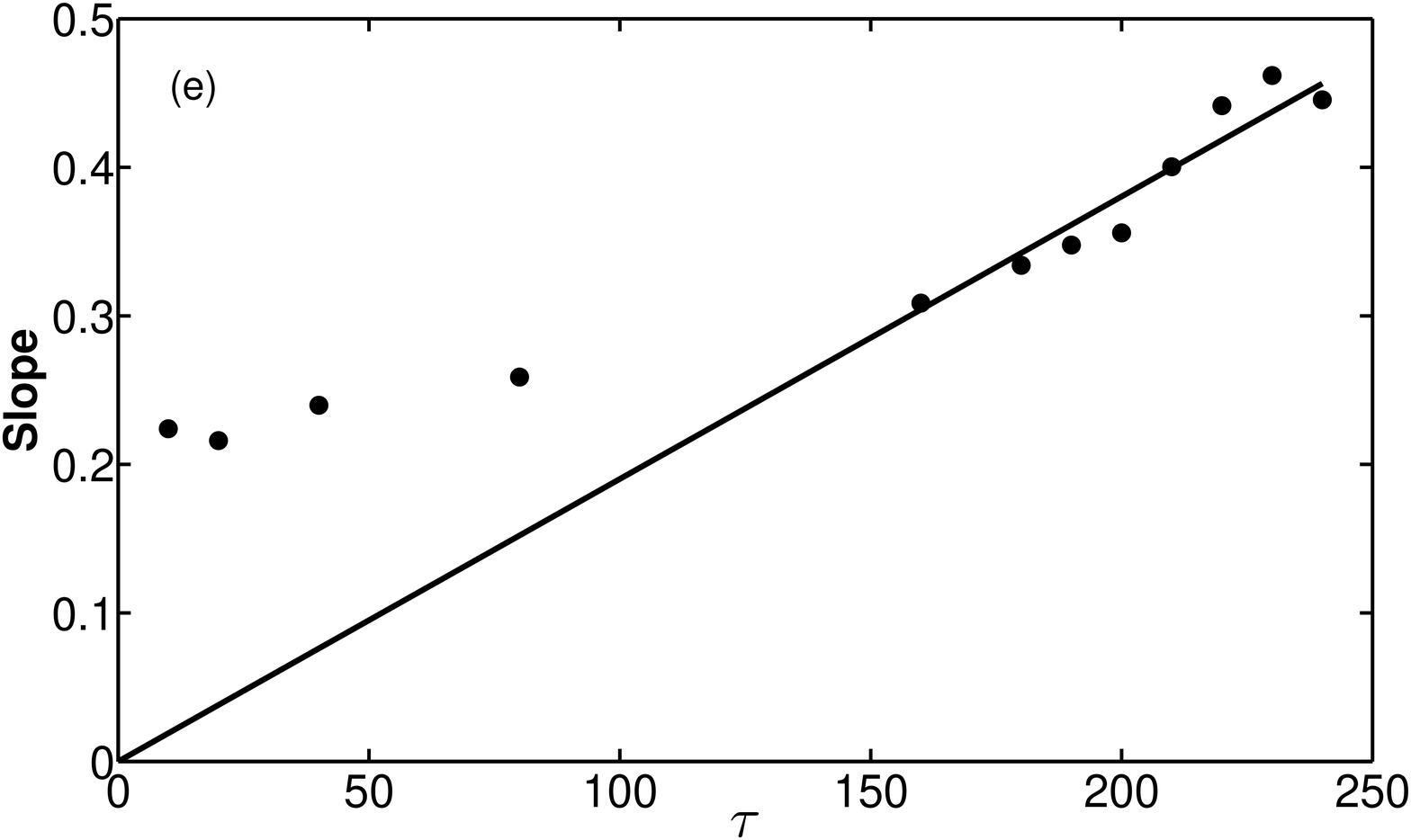}
\caption{\label{fig5} 
Logarithmic probability ratio of positive to 
negative entropy production rates
for $\tau = 10$h (a), $\tau = 80$h (b), $\tau = 160$h (c) and $\tau = 240$h (d) in T21L5 resolution. The slopes of the logarithmic probability ratios against $\tau$ (e).}
\end{figure}

Motivated by the FT, we further analyze the frequency of 
negative $\hat{\lambda}$ events.
The FT relates the ratio of positive to negative 
entropy production rates $\Sigma$
with the mean entropy production rate $\bar{\Sigma}$ and the growth time $\tau$ for sufficiently large $\tau$
\begin{equation} \label{ppstau}
\frac{P(\Sigma=+a)}{P(\Sigma=-a)}
= e^{a \tau \bar{\Sigma}}
\end{equation}
To validate this relation, we plot the logarithm of $P(\hat{\lambda}=a)/P(\hat{\lambda}=-a)$,
the ratio of the probabilities of observing $\hat{\lambda}$ with values $a$ and $-a$.
Figure \ref{fig5} shows this ratio
for different growth times  between 
$\tau=10$h and $\tau=240$h as well as the slopes of the logarithmic probability ratios.
These logarithmic ratios are clear linear functions of 
$\Sigma$. 
Note that \eqref{ppstau} follows for fixed growth time $\tau$
when the fluctuations are Gaussian.
The slopes of the logarithmic probability ratios increase linearly with growth time $\tau$ for $\tau > 160$h as indicated by the relation \eqref{ppstau}.
For finite times there is a non-negligible probability
for negative entropy production rates, 
which vanishes in the long time limit.


\section{Summary and Conclusions}   \label{Sec_SumCon}

The dynamic atmospheric model PUMA 
(Portable University Model of the Atmosphere)
is used to assess the variability of predictability. 
The model has variable vertical and horizontal 
resolutions and is used here for $L=3$ to 20 levels and
$\ell=15$ to $42$ spherical wave numbers.
The model describes a standard dynamical core
with diabatic heating and friction.
In contrast to complex models, 
forcing and dissipation in PUMA are implemented as
linear relaxation processes. 
This keeps the system simple 
to concentrate on the effects of the nonlinear
dynamics. 

The local (time-dependent) largest Lyapunov exponent $\hat{\lambda}$
is estimated by the divergence of two close 200 years 
trajectories,
a reference and a perturbation; the latter
being rescaled in regular time intervals.
The distances between both trajectories are
measured by a Euclidean metric comprising all dynamic variables
at all grid points.

The main interest is in the frequency of negative $\hat{\lambda}$
which is associated with increased predictability
(return of skill) and negative entropy production.
Note that in contrast to the analysis of \cite{patil01}
where a regional analysis is presented, the results
in the present paper correspond to the global circulation.

For moderate resolutions below $\ell=42$ wave numbers
the distribution of $\hat{\lambda}$ shows a considerable ratio
of negative values (up to $20\%$), hence negative entropy production
and violation of the second law of thermodynamics.
This effect is negligible for $\ell \geq 42$. 

The increase of vertical and horizontal resolution
shows different effects: The vertical resolution 
mainly increases the variance of the $\hat{\lambda}$ distribution
while the horizontal resolution primarily increases the mean.
These results suggest a method to determine
an optimal combination of horizontal wave numbers 
and vertical levels in numerical models.
The widening of the distributions for increasing 
vertical resolutions enhances fluctuations
and can lead to an increased frequency of circulation
extremes.

This study demonstrates the validity of the fluctuation theorem in an atmospheric model whith the entropy production rate approximated by the local largest Lyapunov exponent.

The resolutions used here are standard in many
global warming scenarios and even lower
resolutions are used for long term integrations.
The results show that numerical models
used in geophysical fluid dynamics
may be considered as mesoscopic systems.

\begin{acknowledgments}
We like to thank Edilbert Kirk and Davide Faranda 
for helpful discussions.
BS acknowledges funding by the cluster of excellence 
Integrated Climate System Analysis and Prediction (CliSAP,
University of Hamburg). KF 
acknowledges support by a Max-Planck Fellowship.
\end{acknowledgments}

\bibliography{Schalge-FT}

\providecommand{\noopsort}[1]{}\providecommand{\singleletter}[1]{#1}%
\begin{thebibliography}{10}%
\makeatletter
\providecommand \@ifxundefined [1]{%
 \ifx #1\undefined \expandafter \@firstoftwo
 \else \expandafter \@secondoftwo
\fi
}%
\providecommand \@ifnum [1]{%
 \ifnum #1\expandafter \@firstoftwo
 \else \expandafter \@secondoftwo
\fi
}%
\providecommand \enquote [1]{``#1''}%
\providecommand \bibnamefont  [1]{#1}%
\providecommand \bibfnamefont [1]{#1}%
\providecommand \citenamefont [1]{#1}%
\providecommand\href[0]{\@sanitize\@href}%
\providecommand\@href[1]{\endgroup\@@startlink{#1}\endgroup\@@href}%
\providecommand\@@href[1]{#1\@@endlink}%
\providecommand \@sanitize [0]{\begingroup\catcode`\&12\catcode`\#12\relax}%
\@ifxundefined \pdfoutput {\@firstoftwo}{%
 \@ifnum{\z@=\pdfoutput}{\@firstoftwo}{\@secondoftwo}%
}{%
 \providecommand\@@startlink[1]{\leavevmode\special{html:<a href="#1">}}%
 \providecommand\@@endlink[0]{\special{html:</a>}}%
}{%
 \providecommand\@@startlink[1]{%
  \leavevmode
  \pdfstartlink
   attr{/Border[0 0 1 ]/H/I/C[0 1 1]}%
   user{/Subtype/Link/A<</Type/Action/S/URI/URI(#1)>>}%
  \relax
 }%
 \providecommand\@@endlink[0]{\pdfendlink}%
}%
\providecommand \url  [0]{\begingroup\@sanitize \@url }%
\providecommand \@url [1]{\endgroup\@href {#1}{\urlprefix}}%
\providecommand \urlprefix [0]{URL }%
\providecommand \Eprint[0]{\href }%
\@ifxundefined \urlstyle {%
  \providecommand \doi [1]{doi:\discretionary{}{}{}#1}%
}{%
  \providecommand \doi [0]{doi:\discretionary{}{}{}\begingroup
  \urlstyle{rm}\Url }%
}%
\providecommand \doibase [0]{http://dx.doi.org/}%
\providecommand \Doi[1]{\href{\doibase#1}}%
\providecommand \bibAnnote [3]{%
  \BibitemShut{#1}%
  \begin{quotation}\noindent
    \textsc{Key:}\ #2\\\textsc{Annotation:}\ #3%
  \end{quotation}%
}%
\providecommand \bibAnnoteFile [2]{%
  \IfFileExists{#2}{\bibAnnote {#1} {#2} {\input{#2}}}{}%
}%
\providecommand \typeout [0]{\immediate \write \m@ne }%
\providecommand \selectlanguage [0]{\@gobble}%
\providecommand \bibinfo [0]{\@secondoftwo}%
\providecommand \bibfield [0]{\@secondoftwo}%
\providecommand \translation [1]{[#1]}%
\providecommand \BibitemOpen[0]{}%
\providecommand \bibitemStop [0]{}%
\providecommand \bibitemNoStop [0]{.\EOS\space}%
\providecommand \EOS [0]{\spacefactor3000\relax}%
\providecommand \BibitemShut [1]{\csname bibitem#1\endcsname}%
\bibitem{evans93}%
  \BibitemOpen
  \bibfield{author}{%
  \bibinfo {author} {\bibfnamefont{D.~J.}\ \bibnamefont{Evans}}, \bibinfo
  {author} {\bibfnamefont{E.~G.~D.}\ \bibnamefont{Cohen}},\ and\ \bibinfo
  {author} {\bibfnamefont{G.~P.}\ \bibnamefont{Morriss}},\ }%
  \bibfield{journal}{%
  \bibinfo {journal} {Phys.\ Rev.\ Lett.}\ }%
  \textbf{\bibinfo {volume} {71}},\ \bibinfo {pages} {2401} (\bibinfo {year}
  {1993})%
  \bibAnnoteFile{NoStop}{evans93}%
\bibitem{gallavotti95}%
  \BibitemOpen
  \bibfield{author}{%
  \bibinfo {author} {\bibfnamefont{G.}~\bibnamefont{Gallavotti}}\ and\ \bibinfo
  {author} {\bibfnamefont{E.~G.~D.}\ \bibnamefont{Cohen}},\ }%
  \bibfield{journal}{%
  \bibinfo {journal} {Physical Review Letters}\ }%
  \textbf{\bibinfo {volume} {74}},\ \bibinfo {pages} {2694} (\bibinfo {year}
  {1995})%
  \bibAnnoteFile{NoStop}{gallavotti95}%
\bibitem{ciliberto98}%
  \BibitemOpen
  \bibfield{author}{%
  \bibinfo {author} {\bibfnamefont{S.}~\bibnamefont{Ciliberto}}\ and\ \bibinfo
  {author} {\bibfnamefont{C.}~\bibnamefont{Laroche}},\ }%
  \bibfield{journal}{%
  \bibinfo {journal} {Le Journal de Physique {IV}}\ }%
  \textbf{\bibinfo {volume} {08}},\ \bibinfo {pages} {5} (\bibinfo {year}
  {1998})%
  \bibAnnoteFile{NoStop}{ciliberto98}%
\bibitem{gallavotti04}%
  \BibitemOpen
  \bibfield{author}{%
  \bibinfo {author} {\bibfnamefont{G.}~\bibnamefont{Gallavotti}}, \bibinfo
  {author} {\bibfnamefont{L.}~\bibnamefont{Rondoni}},\ and\ \bibinfo {author}
  {\bibfnamefont{E.}~\bibnamefont{Segre}},\ }%
  \bibfield{journal}{%
  \bibinfo {journal} {Physica D: Nonlinear Phenomena}\ }%
  \textbf{\bibinfo {volume} {187}},\ \bibinfo {pages} {338} (\bibinfo {year}
  {2004})%
  \bibAnnoteFile{NoStop}{gallavotti04}%
\bibitem{anderson94}%
  \BibitemOpen
  \bibfield{author}{%
  \bibinfo {author} {\bibfnamefont{J.~L.}\ \bibnamefont{Anderson}}\ and\
  \bibinfo {author} {\bibfnamefont{H.~M.}\ \bibnamefont{van~den Dool}},\ }%
  \bibfield{journal}{%
  \bibinfo {journal} {Mon.\ Wea.\ Rev.}\ }%
  \textbf{\bibinfo {volume} {122}},\ \bibinfo {pages} {507} (\bibinfo {year}
  {1994})%
  \bibAnnoteFile{NoStop}{anderson94}%
\bibitem{rhines75}%
  \BibitemOpen
  \bibfield{author}{%
  \bibinfo {author} {\bibfnamefont{P.~B.}\ \bibnamefont{Rhines}},\ }%
  \bibfield{journal}{%
  \bibinfo {journal} {J.\ Fluid\ Mech.}\ }%
  \textbf{\bibinfo {volume} {69}},\ \bibinfo {pages} {417} (\bibinfo {year}
  {1975})%
  \bibAnnoteFile{NoStop}{rhines75}%
\bibitem{casartelli76}%
  \BibitemOpen
  \bibfield{author}{%
  \bibinfo {author} {\bibfnamefont{M.}~\bibnamefont{Casartelli}}, \bibinfo
  {author} {\bibfnamefont{E.}~\bibnamefont{Diana}}, \bibinfo {author}
  {\bibfnamefont{L.}~\bibnamefont{Galgani}},\ and\ \bibinfo {author}
  {\bibfnamefont{A.}~\bibnamefont{Scotti}},\ }%
  \bibfield{journal}{%
  \bibinfo {journal} {Phys.\ Rev.\ A}\ }%
  \textbf{\bibinfo {volume} {13}},\ \bibinfo {pages} {1921} (\bibinfo {year}
  {1976})%
  \bibAnnoteFile{NoStop}{casartelli76}%
\bibitem{benettin76}%
  \BibitemOpen
  \bibfield{author}{%
  \bibinfo {author} {\bibfnamefont{G.}~\bibnamefont{Benettin}}, \bibinfo
  {author} {\bibfnamefont{L.}~\bibnamefont{Galgani}},\ and\ \bibinfo {author}
  {\bibfnamefont{J.-M.}\ \bibnamefont{Strelcyn}},\ }%
  \bibfield{journal}{%
  \bibinfo {journal} {Phys.\ Rev.\ A}\ }%
  \textbf{\bibinfo {volume} {14}},\ \bibinfo {pages} {2338} (\bibinfo {year}
  {1976})%
  \bibAnnoteFile{NoStop}{benettin76}%
\bibitem{young03}%
  \BibitemOpen
  \bibfield{author}{%
  \bibinfo {author} {\bibfnamefont{L.-S.}\ \bibnamefont{Young}},\ }%
  \emph{\bibinfo {title} {Entropy}}\ (\bibinfo {publisher} {Princeton
  University Press},\ \bibinfo {year} {2003})\ Chap.~\bibinfo {chapter} {16}%
  \bibAnnoteFile{NoStop}{young03}%
\bibitem{fujisaka83}%
  \BibitemOpen
  \bibfield{author}{%
  \bibinfo {author} {\bibfnamefont{H.}~\bibnamefont{Fujisaka}},\ }%
  \bibfield{journal}{%
  \Doi{10.1143/PTP.70.1264}{\bibinfo {journal} {Progress of Theoretical
  Physics}}\ }%
  \textbf{\bibinfo {volume} {70}},\ \bibinfo {pages} {1264} (\bibinfo {year}
  {1983})%
  \bibAnnoteFile{NoStop}{fujisaka83}%
\bibitem{yoshida89}%
  \BibitemOpen
  \bibfield{author}{%
  \bibinfo {author} {\bibfnamefont{T.}~\bibnamefont{Yoshida}}\ and\ \bibinfo
  {author} {\bibfnamefont{S.}~\bibnamefont{Miyazaki}},\ }%
  \bibfield{journal}{%
  \Doi{10.1143/PTPS.99.64}{\bibinfo {journal} {Progress of Theoretical Physics
  Supplement}}\ }%
  \textbf{\bibinfo {volume} {99}},\ \bibinfo {pages} {64} (\bibinfo {year}
  {1989})%
  \bibAnnoteFile{NoStop}{yoshida89}%
\bibitem{fraedrich05}%
  \BibitemOpen
  \bibfield{author}{%
  \bibinfo {author} {\bibfnamefont{K.}~\bibnamefont{Fraedrich}}, \bibinfo
  {author} {\bibfnamefont{E.}~\bibnamefont{Kirk}}, \bibinfo {author}
  {\bibfnamefont{U.}~\bibnamefont{Luksch}},\ and\ \bibinfo {author}
  {\bibfnamefont{F.}~\bibnamefont{Lunkeit}},\ }%
  \bibfield{journal}{%
  \bibinfo {journal} {Meteorol.\ Zeitschrift}\ }%
  \textbf{\bibinfo {volume} {14}},\ \bibinfo {pages} {735} (\bibinfo {year}
  {2005})%
  \bibAnnoteFile{NoStop}{fraedrich05}%
\bibitem{fraedrich12}%
  \BibitemOpen
  \bibfield{author}{%
  \bibinfo {author} {\bibfnamefont{K.}~\bibnamefont{Fraedrich}},\ }%
  \bibfield{journal}{%
  \bibinfo {journal} {Eur.\ Phys.\ J.\ Plus}\ }%
  \textbf{\bibinfo {volume} {127}},\ \bibinfo {pages} {53} (\bibinfo {year}
  {2012})%
  \bibAnnoteFile{NoStop}{fraedrich12}%
\bibitem{orszag70}%
  \BibitemOpen
  \bibfield{author}{%
  \bibinfo {author} {\bibfnamefont{S.~A.}\ \bibnamefont{Orszag}},\ }%
  \bibfield{journal}{%
  \bibinfo {journal} {J.\ Atmos.\ Sci.}\ }%
  \textbf{\bibinfo {volume} {27}},\ \bibinfo {pages} {890} (\bibinfo {year}
  {1970})%
  \bibAnnoteFile{NoStop}{orszag70}%
\bibitem{harlim05}%
  \BibitemOpen
  \bibfield{author}{%
  \bibinfo {author} {\bibfnamefont{J.}~\bibnamefont{Harlim}}, \bibinfo {author}
  {\bibfnamefont{M.}~\bibnamefont{Oczkowski}}, \bibinfo {author}
  {\bibfnamefont{J.~A.}\ \bibnamefont{Yorke}}, \bibinfo {author}
  {\bibfnamefont{E.}~\bibnamefont{Kalnay}},\ and\ \bibinfo {author}
  {\bibfnamefont{B.~R.}\ \bibnamefont{Hunt}},\ }%
  \bibfield{journal}{%
  \bibinfo {journal} {Phys.\ Rev.\ Lett.}\ }%
  \textbf{\bibinfo {volume} {94}},\ \bibinfo {pages} {228501} (\bibinfo {year}
  {2005})%
  \bibAnnoteFile{NoStop}{harlim05}%
\bibitem{smith99}%
  \BibitemOpen
  \bibfield{author}{%
  \bibinfo {author} {\bibfnamefont{L.~A.}\ \bibnamefont{Smith}}, \bibinfo
  {author} {\bibfnamefont{C.}~\bibnamefont{Ziehmann}},\ and\ \bibinfo {author}
  {\bibfnamefont{K.}~\bibnamefont{Fraedrich}},\ }%
  \bibfield{journal}{%
  \bibinfo {journal} {Quart.\ J.\ Roy.\ Meteorol.\ Soc.}\ }%
  \textbf{\bibinfo {volume} {125}},\ \bibinfo {pages} {2855} (\bibinfo {year}
  {1999})%
  \bibAnnoteFile{NoStop}{smith99}%
\bibitem{peixoto92}%
  \BibitemOpen
  \bibfield{author}{%
  \bibinfo {author} {\bibfnamefont{J.~P.}\ \bibnamefont{Peixoto}}\ and\
  \bibinfo {author} {\bibfnamefont{A.~H.}\ \bibnamefont{Oort}},\ }%
  \emph{\bibinfo {title} {Physics of climate}}\ (\bibinfo {publisher} {AIP, New
  York},\ \bibinfo {year} {1992})%
  \bibAnnoteFile{NoStop}{peixoto92}%
\bibitem{patil01}%
  \BibitemOpen
  \bibfield{author}{%
  \bibinfo {author} {\bibfnamefont{D.~J.}\ \bibnamefont{Patil}}, \bibinfo
  {author} {\bibfnamefont{B.~R.}\ \bibnamefont{Hunt}}, \bibinfo {author}
  {\bibfnamefont{E.}~\bibnamefont{Kalnay}}, \bibinfo {author}
  {\bibfnamefont{J.~A.}\ \bibnamefont{Yorke}},\ and\ \bibinfo {author}
  {\bibfnamefont{E.}~\bibnamefont{Ott}},\ }%
  \bibfield{journal}{%
  \bibinfo {journal} {Phys.\ Rev.\ Lett.}\ }%
  \textbf{\bibinfo {volume} {86}},\ \bibinfo {pages} {5878} (\bibinfo {year}
  {2001})%
  \bibAnnoteFile{NoStop}{patil01}%
\end{thebibliography}%

\end{document}